\documentclass[amsfonts, amssymb, amsmath, aps, 10pt, pra, twocolumn, english, floatfix, showkeys]{revtex4-2}

\usepackage[utf8]{inputenc}
\usepackage[english]{babel}
\usepackage[T1]{fontenc}
\usepackage{graphicx}
\usepackage{bm}
\usepackage[dvipsnames]{xcolor}
\usepackage{physics}
\usepackage{dsfont}
\usepackage{mathtools}
\usepackage[nolist]{acronym}
\usepackage{quantikz}
\usepackage{subcaption}
\usepackage{orcidlink}
\usepackage{hyperref}

\definecolor{colorblue}{HTML}{0000FF}
\definecolor{colorgreen}{HTML}{008000}
\definecolor{colorred}{HTML}{ff0100}
\definecolor{colororange}{HTML}{ffa500}

\newsavebox{\boxred}
\savebox{\boxred}{\tikz \fill[color=colorred] (0,0) rectangle (2ex,2ex);}

\newsavebox{\orangecircle}
\savebox{\orangecircle}{\tikz \filldraw[color=colororange, fill=colororange, line width=0.5mm] (0,0) circle (1ex);}

\newsavebox{\bluediamond}
\savebox{\bluediamond}{%
    \tikz \filldraw[color=colorblue, fill=colorblue, line width=0.5mm]
    (0,1ex) -- (1ex,0) -- (0,-1ex) -- (-1ex,0) -- cycle;
}

\newsavebox{\greentriangle}
\savebox{\greentriangle}{%
    \tikz \fill[color=colorgreen]
    (0,1ex) -- (-1ex,-0.5ex) -- (1ex,-0.5ex) -- cycle;
}

\newcommand{\cH}{\mathcal{H}}

\newcommand{\cP}{\mathcal{P}}

\graphicspath{{figures/}}

\hypersetup{
colorlinks = true,
linkcolor = RoyalBlue,
citecolor = RoyalBlue,
urlcolor = RoyalBlue
}

\begin{document}

\title{Structured Parameterization and Non-Stabilizerness in Hypergraph QAOA}

\author{Evan Camilleri \orcidlink{0009-0009-5398-2515}}
\email{evan.camilleri.22@um.edu.mt}
\author{Andr\'e Xuereb \orcidlink{0000-0002-9030-7334}}
\author{Tony J. G. Apollaro \orcidlink{0000-0002-9324-9336}}
\author{Mirko Consiglio \orcidlink{0000-0001-8730-0206}}
\affiliation{Department of Physics, University of Malta, Msida MSD 2080, Malta}

\date{\today}

\begin{abstract}
    The \ac{QAOA} has emerged as a promising candidate for demonstrating quantum advantage on \ac{NISQ} devices. While various \ac{QAOA} parameterization schemes exist, ranging from the original single-angle approach to the more expressive \ac{MA-QAOA} and \ac{AA-QAOA}, each presents distinct trade-offs between expressiveness and classical optimization complexity. In this work, we introduce the \ac{kA-QAOA}, a parameterization scheme that groups cost function terms by their $k$-body interaction order, providing a natural middle ground between parameter efficiency and solution quality. This approach is particularly well-suited for combinatorial optimization problems defined on hypergraphs, where multi-body interactions naturally arise in applications such as Boolean satisfiability and resource allocation with multi-party constraints. We benchmark \ac{kA-QAOA} against standard \ac{SA-QAOA}, \ac{MA-QAOA}, and \ac{AA-QAOA} on two problem classes: 3-uniform cyclic sign-alternating hypergraphs and random coefficient hypergraphs. Our results demonstrate that \ac{kA-QAOA} achieves approximation ratios comparable to \ac{MA-QAOA} while requiring significantly fewer function evaluations, thereby reducing quantum resource consumption.
\end{abstract}

\keywords{Quantum Approximate Optimization Algorithm, Quantum Magic, Variational Quantum Algorithms, Combinatorial Optimization}

\maketitle

\begin{acronym}
    \acro{QUBO}[QUBO]{quadratic unconstrained binary optimization}
    \acro{PUBO}[PUBO]{polynomial unconstrained binary optimization}
    \acro{QAOA}[QAOA]{quantum approximate optimization algorithm}
    \acro{MA-QAOA}[MA-QAOA]{multi-angle quantum approximate optimization algorithm}
    \acro{AA-QAOA}[AA-QAOA]{automorphic-angle quantum approximate optimization algorithm}
    \acro{kA-QAOA}[$k$A-QAOA]{$k$-interaction-angle quantum approximate optimization algorithm}
    \acro{SA-QAOA}[SA-QAOA]{single-angle quantum approximate optimization algorithm}
    \acro{PQC}[PQC]{parameterized quantum circuit}
    \acro{VQA}[VQA]{variational quantum algorithm}
    \acro{SRE}[SRE]{stabilizer R\'enyi entropy}
    \acro{NISQ}[NISQ]{noisy intermediate-scale quantum}
    \acro{AR}[AR]{approximation ratio}
    \acro{TQA}[TQA]{trotterized quantum annealing}
    \acro{BFGS}[BFGS]{Broyden--Fletcher--Goldfarb--Shanno}
    \acro{JSSP}[JSSP]{job-shop scheduling problem}
    \acro{nfev}[nfev]{number of function evaluations}
\end{acronym}

\section{Introduction}

The advent of quantum technologies \cite{Ezratty2025} has opened new avenues for tackling combinatorial optimization problems \cite{Toledo1999, Lucas2014, Yarkoni2022, Wang2023}, offering the potential to accelerate and enhance solutions beyond what classical algorithms can achieve. While fully fault-tolerant quantum computers remain a long-term goal, hybrid quantum--classical algorithms operating in the \ac{NISQ} \cite{Preskill2018} era have emerged as promising candidates for demonstrating practical quantum advantage in the near term. Among these approaches, the \ac{QAOA} \cite{Farhi2014} has attracted considerable attention as a \ac{VQA} that bridges quantum adiabatic optimization principles with gate-based quantum computation.

Understanding the computational resources that enable quantum algorithms to outperform their classical counterparts has become a central question in quantum information science. Initially, quantum entanglement \cite{Amico2008, Eisert2010} was widely considered the fundamental resource underlying quantum advantage \cite{Preskill2012}. However, this perspective proved incomplete. The existence of highly entangled states that admit efficient classical simulation, most notably stabilizer states, which can be efficiently simulated via Clifford circuits as established by the Gottesman--Knill theorem \cite{Aaronson2004}, demonstrates that entanglement alone cannot explain quantum computational power.

This observation has motivated the study of nonstabilizerness \cite{Bravyi2005}, a quantum resource that quantifies how far a quantum state deviates from the set of stabilizer states. This property, commonly referred to as quantum magic or simply magic, has emerged as a crucial ingredient for quantum advantage. Unlike entanglement, magic captures the computational complexity introduced by non-Clifford operations, which are necessary to achieve universal quantum computation.

The interplay between magic, entanglement, and algorithmic performance in \ac{QAOA} remains an active area of investigation. Recent studies suggest that the generation and utilization of magic resources may be intimately connected to the algorithm's ability to solve hard optimization problems. Understanding how magic accumulates throughout \ac{QAOA} circuits, how it correlates with solution quality, and whether it can serve as a diagnostic tool for quantum advantage represents an important frontier in quantum algorithm analysis. This work aims to investigate these connections systematically, providing insights into the role of magic as a computational resource in variational quantum optimization.

Many combinatorial optimization problems of practical interest can be naturally formulated on graphs, where binary variables correspond to vertices and interactions appear as edges. However, numerous real-world optimization scenarios involve interactions among three or more entities simultaneously. For instance, Boolean satisfiability (SAT) problems involve clauses that constrain multiple variables simultaneously, where each clause represents a higher-order relationship that cannot be decomposed into pairwise interactions without losing essential structure \cite{Feige2006}. Similarly, resource allocation problems with multi-party constraints and scheduling problems with group dependencies, such as \ac{JSSP}, require modeling simultaneous interactions among multiple entities \cite{Heydaribeni2024, Singh2025}. Such constraints cannot be fully captured by pairwise (quadratic) penalties alone except if converted to a quadratic formulation \cite{Dattani2019}, but this requires introducing ancilla variables and additional penalty terms, which significantly increase the problem size. Hypergraphs provide a natural framework for representing these multi-body interactions, where hyperedges connect arbitrary subsets of vertices \cite{Guruswami2010, Farhi2014}.

This paper is organized as follows: In Sec. \ref{sec:QAOA}, we introduce the \ac{QAOA} and its various formulations, including \ac{SA-QAOA}, \ac{MA-QAOA} and \ac{AA-QAOA}, before presenting our proposed \ac{kA-QAOA} variant, as well as reviewing the stabilizer R\'enyi entropy as a measure of quantum magic and its role in characterizing nonstabilizerness. Section \ref{sec:Optimization} discusses the optimization details, such as the choice of classical optimizer and parameter initialization strategies. In Sec. \ref{sec:Results}, we present our results comparing \ac{kA-QAOA} against \ac{SA-QAOA}, \ac{MA-QAOA}, and \ac{AA-QAOA} across two benchmark problem classes: 3-uniform cyclic sign-alternating hypergraphs and random coefficient hypergraphs. We evaluate performance using approximation ratio, fidelity, and the \ac{nfev} required for convergence. We conclude in Sec. \ref{sec:Conclusion} with a discussion of the implications of our results for NISQ-era quantum optimization algorithms. Appendices \ref{app:3-uniform_results} and \ref{app:random_coefficient_results} present the results discussed in Sec. \ref{sec:Results} in a graphical format.

\section{QAOA} \label{sec:QAOA}

The \ac{QAOA} \cite{Farhi2014, Giovagnoli2025} is a \ac{VQA} \cite{Cerezo2021} designed to produce approximate solutions for combinatorial problems that are difficult to solve using classical approaches. The \ac{QAOA} falls under the classes of hybrid quantum-classical algorithms, whereby a \ac{PQC} is trained on a quantum computer via a classical optimizer. 

The $n$-qubit generic \ac{QAOA}, which we refer to as \ac{SA-QAOA}, consists of initially preparing the ground state of a known, simple Hamiltonian called the mixer Hamiltonian, such as $\cH = \sum_i \sigma^x_i$, with the ground state being $\ket{+}^{\otimes n}$, which can easily be prepared via a layer of Hadamard gates.

The \textit{Ansatz} then consists of applying alternating layers of `cost' and `mixer' \acp{PQC}. The cost \ac{PQC} is typically made up of parameterized operators found in the cost Hamiltonian $\cH_C$, with the cost function related to finding its ground state, such that $U_C(\gamma) = \exp(-\imath \gamma \cH_C)$. The `mixer' layer similarly consists of parameterized operators of the mixer Hamiltonian $\cH_M$, which generally contains solely $U_M(\beta) = \sum_i R_x(\beta)$ gates. Both $\gamma$ and $\beta$ are variational parameters of their respective layer.

Each of these layers is typically repeated for a fixed number of times, $p$, such that the limit $p \to \infty$ achieves the minimization of the cost function. In the case of a \ac{QUBO} problem, the classical cost function is typically constructed as follows
\begin{equation}
C(\bm{x}) = \bm{x}^T\bm{Q}\bm{x} + \bm{q}^T\bm{x} = \sum_{i, j = 1}^n Q_{ij} x_i x_j + \sum_{i=1}^n q_i x_i, 
\end{equation}
where $\bm{Q}$ is a real symmetric matrix, and $\bm{x}$ are $n$-bit strings. The classical optimizer finds the optimal bit string
\begin{equation}
\bm{x}^* = \underset{\bm{x} \in \{0, 1\}^n}{\arg \min}~C(\bm{x}).
\end{equation}

\ac{QUBO} problems can be transformed to Ising problems by applying the transformation $x_i \to (1 - \sigma^z_i) / 2$, such that the cost function then becomes
\begin{align}
C(\bm{z}) &= \sum_{i, j = 1}^n Q_{ij} x_i x_j + \sum_{i=1}^n q_i x_i \nonumber \\
&= \sum_{i, j = 1}^n \frac{Q_{ij}}{4} (1 - \sigma^z_i)(1 - \sigma^z_j) + \sum_{i=1}^n \frac{q_i}{2} (1 - \sigma^z_i) \nonumber \\
&= \sum_{i, j = 1}^n \frac{Q_{ij}}{4} \sigma^z_i\sigma^z_j - \sum_{i=1}^n \frac{1}{2} \left(q_i + \sum_{j=1}^n Q_{ij}\right)\sigma^z_i,
\end{align}
(removing the emergent constant term) which is equivalent to an Ising model where $J_{ij} \equiv -Q_{ij} / 4$ and $h_i \equiv (q_i + \sum_j Q_{ij}) / 2$, such that
\begin{equation}
C(\bm{z}) = -\sum_{i, j = 1}^n J_{ij} \sigma^z_i\sigma^z_j - \sum_{i=1}^n h_i \sigma^z_i.
\end{equation}
This means that finding the solution of a \ac{QUBO} problem is equivalent to finding the ground state of the Ising model.

\subsection{Variants of QAOA}

In the original formulation of \ac{SA-QAOA} by \citet{Farhi2014}, each cost and mixer layer consists of an individual single parameter $\gamma_i, \beta_i,\forall i \in [p]$, for a total of $2p$ parameters. \citet{Herrman2022} proposed a \ac{MA-QAOA} approach, where each individual parameterized gate in each layer consists of its own separate angle.

This approach requires parameters for the classical optimizer, implying that the classical algorithm will generally require more iterations to converge, with the prevalence of barren plateaus \cite{McClean2018, Cerezo2021, Larocca2025} increasing. On the other hand, due to the more expressive power \cite{Bharti2022} of \ac{MA-QAOA}, the number of layers will also generally be less than the number of layers needed for \ac{SA-QAOA} to achieve the same fidelity. Note that \ac{SA-QAOA} is a special case of \ac{MA-QAOA}.

As a trade-off between the \ac{SA-QAOA} proposed and \ac{MA-QAOA}, \citet{Shi2022} proposed best-1sym-\ac{QAOA}, max-sym-\ac{QAOA}, and randgroup-\ac{QAOA} in which they exploit the symmetries of the underlying graphs of the problems being studied, which we compactly refer to as \ac{AA-QAOA}. A common angle is used across all vertices falling under the same symmetry group, thus reducing the number of angles required for \ac{MA-QAOA}. If there are no symmetries in the graph then \ac{AA-QAOA} reduces to \ac{MA-QAOA}.

\begin{figure*}[t]
\centering
\scalebox{0.9}{%
\begin{quantikz} 
\lstick{$q_0$} & \gate{H} & \gate{R_{Z}(\gamma_1)}\gategroup[1, steps=1, style={dashed, rounded corners, fill=blue!20, inner xsep=2pt},background, label style={label position=above, yshift=0.2cm}]{$R_{z_0}$ Gate} & \ctrl{1}\gategroup[2, steps=3, style={dashed, rounded corners, fill=green!20, inner xsep=2pt},background, label style={label position=above, yshift=0.2cm}]{$R_{{z_0}{z_1}}$ Gate} & \qw & \ctrl{1} & \ctrl{2}\gategroup[3, steps=3, style={dashed, rounded corners, fill=green!20, inner xsep=2pt},background, label style={label position=above, yshift=0.2cm}]{$R_{{z_0}{z_2}}$ Gate} & \qw & \ctrl{2} & \qw & \qw & \qw & \qw \\
\lstick{$q_1$} & \gate{H} & \qw & \targ{} & \gate{R_z(\gamma_2)} & \targ{} & \qw & \qw & \qw & \ctrl{2}\gategroup[3, steps=3, style={dashed, rounded corners, fill=green!20, inner xsep=2pt},background, label style={label position=below, yshift=-0.6cm}]{$R_{{z_1}{z_3}}$ Gate} & \qw & \ctrl{2} & \qw \\
\lstick{$q_2$} & \gate{H} & \qw & \ctrl{1}\gategroup[2, steps=3, style={dashed, rounded corners, fill=green!20, inner xsep=2pt},background, label style={label position=below, yshift=-0.6cm}]{$R_{{z_2}{z_3}}$ Gate} & \qw & \ctrl{1} & \targ{} & \gate{R_z(\gamma_2)} & \targ{} & \qw & \qw & \qw & \qw \\
\lstick{$q_3$} & \gate{H} & \gate{R_{Z}(\gamma_1)}\gategroup[1, steps=1, style={dashed, rounded corners, fill=blue!20, inner xsep=2pt},background, label style={label position=below, yshift=-0.6cm}]{$R_{z_3}$ Gate} & \targ{} & \gate{R_z(\gamma_2)} & \targ{} & \qw & \qw & \qw & \targ{} & \gate{R_z(\gamma_2)} & \targ{} & \qw \\
	\\
	\\
\lstick{$q_0$} & \ctrl{1}\gategroup[3, steps=5, style={dashed, rounded corners, fill=yellow!20, inner xsep=2pt},background, label style={label position=below, yshift=-1.9cm}]{$R_{z_0z_1z_2}$ Gate} & \qw & \qw & \qw & \ctrl{1} && \qw & \qw & \qw & \qw & \gate{R_x(\beta_1)}\gategroup[1, steps=1, style={dashed, rounded corners, fill=red!20, inner xsep=2pt},background]{} & \qw \\
\lstick{$q_1$} & \targ{} & \ctrl{1} & \qw & \ctrl{1} & \targ{} & \ctrl{1}\gategroup[3, steps=5, style={dashed, rounded corners, fill=yellow!20, inner xsep=2pt},background, label style={label position=below, yshift=-0.7cm}]{$R_{z_1z_2z_3}$ Gate} & \qw & \qw & \qw & \ctrl{1} & \gate{R_x(\beta_2)}\gategroup[1, steps=1, style={dashed, rounded corners, fill=red!20, inner xsep=2pt},background]{} & \qw \\
\lstick{$q_2$} & \qw & \targ{} & \gate{R_z(\gamma_3)} & \targ{} & \qw & \targ{} & \ctrl{1} & \qw & \ctrl{1} & \targ{} & \gate{R_x(\beta_3)}\gategroup[1, steps=1, style={dashed, rounded corners, fill=red!20, inner xsep=2pt},background]{} & \qw \\
\lstick{$q_3$} & \qw & \qw & \qw & \qw & \qw & \qw & \targ{} & \gate{R_z(\gamma_3)} & \targ{} & \qw & \gate{R_x(\beta_4)}\gategroup[1, steps=1, style={dashed, rounded corners, fill=red!20, inner xsep=2pt},background, label style={label position=below, yshift=-0.7cm}]{$R_{x_i}$ Gates} & \qw
\end{quantikz}
}
\caption{\ac{kA-QAOA} four-qubit quantum circuit for Eq. \eqref{eq:c3:quantum_circuit_kaqaoa_cf} with $p=1$ layers having parameterized angles $\gamma_1$ for $R_{z_i}$ (blue) gates, $\gamma_2$ for $R_{z_i z_j}$ (green) gates, $\gamma_3$ for $R_{z_i z_j z_k}$ (yellow) gates, as well as $\beta_1$, $\beta_2$, $\beta_3$, and $\beta_4$ for $R_{x_i}$ (red) gates.}
\label{fig:c3:quantum_circuit_kaqaoa}
\end{figure*}
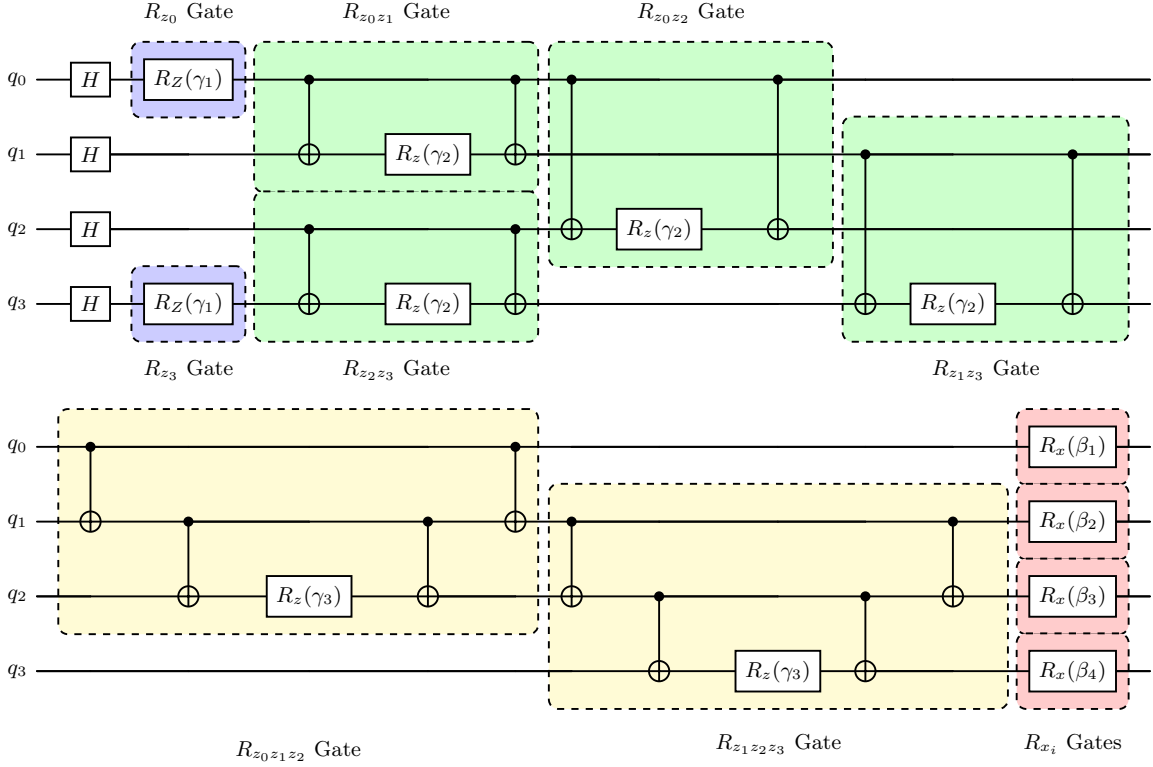

In this manuscript we propose an alternative middle ground approach between the original \ac{SA-QAOA} formulation and \ac{MA-QAOA}, the \ac{kA-QAOA}, which is similarly a special case of \ac{MA-QAOA}. In contrast to \ac{MA-QAOA}, this approach assigns a common angle to all terms in the cost function that involve the same number of interactions between Boolean variables in a \ac{PUBO}. This is particularly useful when there are no symmetry groups in the graph, such as in arbitrarily weighted graphs. As an example, consider a four-qubit Cost Hamiltonian
\begin{align}
C(\bm{z}) &= \sigma^z_0\sigma^z_1\sigma^z_2 - \sigma^z_0\sigma^z_1 - \sigma^z_0\sigma^z_2 + \sigma^z_0 \nonumber \\ & - \sigma^z_1\sigma^z_2\sigma^z_3 + \sigma^z_1\sigma^z_3 + \sigma^z_2\sigma^z_3 - \sigma^z_3,
\label{eq:c3:quantum_circuit_kaqaoa_cf}
\end{align}
which translates to the circuit in Fig. \ref{fig:c3:quantum_circuit_kaqaoa}. The one-qubit terms $\sigma^z_0$ and $\sigma^z_3$ are represented by an $R_z(\gamma_1)$ gate on qubits 1 and 3, as shown by the blue boxes. The two-qubit terms $\sigma^z_0\sigma^z_1$, $\sigma^z_0\sigma^z_2$, $\sigma^z_1\sigma^z_3$ and $\sigma^z_2\sigma^z_3$ are represented by $R_{zz}(\gamma_2)$ gates, as shown by the green boxes. The three-qubit terms $\sigma^z_0\sigma^z_1\sigma^z_2$ and $\sigma^z_1\sigma^z_2\sigma^z_3$ are represented by a $R_{zzz}(\gamma_3)$ gates, as shown by the yellow boxes. Finally, the mixer layer applies an $R_x(\beta_i)$ gate on each qubit with a separate parameter for each qubit. No grouping is carried out for the mixer layer in the \ac{kA-QAOA}.

Our \ac{kA-QAOA} approach is naturally suited for such problems as it partitions cost function terms by their interaction order ($k$-body interactions), directly reflecting the underlying hypergraph structure. This design choice anticipates the trajectory of quantum hardware development. While standard superconducting processors typically decompose $k$-body terms into sequences of two-qubit gates, emerging platforms such as trapped ions and neutral atom arrays increasingly support native multi-qubit operations (e.g., global Mølmer–Sørensen gates) \cite{Stein2023}. \ac{kA-QAOA} is uniquely positioned to exploit these hardware-native operators, bypassing the overhead of quadratization and the necessity for deep \texttt{CNOT} staircases or ancilla qubits characteristic of \ac{QUBO}-reduced \acp{PUBO}. In doing so, \ac{kA-QAOA} attempts to bridge the gap between abstract problem topology and physical hardware capabilities.

\subsection{Quantifying Magic}

The \ac{SRE} characterizes how a pure state $\ket{\psi}$ of $N$ $D$-dimensional qudits spreads over the basis of Pauli strings \cite{Leone2022}
\begin{equation}
    M_\alpha = \frac{1}{1 - \alpha}\log_2\left( \sum_{P \in \cP_N} \frac{|\expval{P}{\psi}|^{2\alpha}}{D^N} \right),
\end{equation}
where $\cP_N$ is the set of $n$-qudit Pauli strings. $M_\alpha \equiv 0$ if and only if $\ket{\psi}$ is a stabilizer state \cite{Gross2021, Haug2023}.

Recent investigations reveal that QAOA exhibits a characteristic ``magic barrier'' during optimization, a transient buildup of nonstabilizerness that occurs even when both initial and target states have zero or low magic \cite{Capecci2025}. \Ac{QAOA} typically begins in a stabilizer state $\ket{+}^{\otimes n}$ and ideally converges to low-magic ground states, yet it must traverse a regime of increased magic to achieve high-fidelity solutions. This non-monotonic behavior where magic initially rises rapidly, reaches a maximum, and then decreases, suggests that intermediate magic generation is necessary for solving hard optimization problems.

The generation and utilization of magic resources in \ac{QAOA} circuits may be intimately connected to the algorithm's ability to outperform classical approaches. Different \ac{QAOA} parameterizations can exhibit varying efficiency in magic consumption, with greater classical optimization freedom potentially leading to unnecessary nonstabilizerness consumption. Understanding how magic accumulates across different \ac{QAOA} variants, particularly our \ac{kA-QAOA} approach where parameters are grouped by interaction order, provides insight into resource-efficient quantum optimization.

\subsection{Job-shop scheduling problem}

The \ac{JSSP} is one of the most complex and challenging NP-complete problems in combinatorial optimization. The primary objective of \ac{JSSP} is to find the most efficient way to schedule a number of jobs and their respective operations while considering several constraints.  Usually, the goal of \ac{JSSP} is to minimize completion time. Still, the cost function can also be designed to maximize profitability or any other metric important to the organization performing the \ac{JSSP} procedure.

The \ac{JSSP} problem deals with a number of machines $m$ available to process $n$ jobs, each with different known due dates. Each job consists of a number of ordered operations, each with known processing times and each assigned to specific machines. The main rules of any \ac{JSSP} algorithm are usually: (1) each machine can process only one operation at any given time; (2) there is a sequence that operations within a job must follow; (3) operations are non-pre-emptable --- once an operation starts, it has to be completed without interruption; (4) tasks from different jobs are independent.

The choice to investigate 3-uniform cyclic sign-alternating hypergraphs and random coefficient hypergraphs in Section \ref{sec:Results}, is motivated by several complementary considerations. First, 3-uniform cyclic hypergraphs represent a computationally hard class of problems with known theoretical complexity bounds \cite{Garey1979}, providing a rigorous testbed for evaluating the effectiveness of \ac{kA-QAOA} against established benchmarks. Their regular structure with genuine three-body interactions ensures that the algorithm's performance can be directly attributed to its handling of $k$-body terms rather than graph-specific artifacts. Second, the random coefficient hypergraphs with $J_i \in \{-1, +1, 0\}$ offer a contrasting scenario that better reflects real-world problem instances, where interactions may be weighted arbitrarily and lack exploitable symmetries. This combination allows us to assess both the best-case performance on structured problems and the robustness on disordered instances.

\section{Optimization Details} \label{sec:Optimization}

In this section, we discuss the optimization details, including our choices of classical optimizers and parameter initialization strategies.

\subsection{Classical Optimizers}

The role of the classical optimizer is to tune the parameters of the \ac{VQA} to optimize the cost function. We initially investigated two different optimizers; \ac{BFGS} \cite{Nocedal2006}, a quasi-Newton method that approximates the Hessian using gradient evaluations; and the Powell method \cite{Powell1964}, which operates using one-dimensional minimizations along a set of search directions. Each minimization run employed the relative default optimizer settings from the \texttt{SciPy} library \cite{Virtanen2020}.

\subsection{Parameter Initialization Techniques}

One particular physically-motivated initialization technique is \ac{TQA} \cite{Sack2021}, which is parameterized by the Trotter time step. \ac{TQA} is based on the adiabatic theorem, and using the equation $\hat{H}(t) = (t/T)\hat{H}_I + (1 - t/T)\hat{H}_C$ for a total number of fixed time steps $T$ at time $t$, the angles are set as 
\begin{equation}
    \gamma_i = \frac{i}{p}\Delta t, \beta_i = \left(1-\frac{i}{p}\right)\Delta t,
\end{equation}
where $i \in [p]$ for the different layers, and time step $\Delta t = T / p$.

In this manuscript, we applied four different parameter initialization techniques: (1) $1/p$, Each layer has all its parameters initialized to $1/p$, where $p$ is the layer number; (2) $R(0, 1/p)$, similar to the previous one but each parameter is initialized to a random number between $0$ and $1/p$; (3) $R(0, 2\pi/p)$, similar to the previous one but each parameter is initialized to a random number between $0$ and $2\pi/p$; (4) \ac{TQA}$_{0.75}$, based on the approach by \citet{Sack2021}, where we set $\Delta t = 0.75$.

\section{Results} \label{sec:Results}

The results discussed are presented here, and the plots shown in Appendices \ref{app:3-uniform_results} and \ref{app:random_coefficient_results}, are taken as the total average between all parameter initialization strategies, as well as both optimization methods, where the error bars represent the standard deviation of all samples. The mixer Hamiltonian is taken to be $\sum_i \sigma^x_i$ in all \ac{QAOA} variants.

To evaluate the success of \ac{QAOA} across different configurations and in comparison to other variants, it is essential to use appropriate figures of merit. This section introduces the \ac{AR} \cite{Farhi2014} and fidelity \cite{Uhlmann1976} metrics.

The \ac{AR} is the figure of merit that measures algorithm efficiency in approximation algorithms, it is defined as
\begin{equation}
    \text{AR} = \frac{C_\text{min}}{C_\text{opt}},
\end{equation}
where $C_\text{min}$ is the minimum attained value of the cost function, and $C_\text{opt}$ is the true optimal value of the cost function (typically computed classically via brute force).

Another figure of merit one can consider is the fidelity of the output state compared to the optimal state. The fidelity of the \ac{QAOA} output state $\ket{\psi}$ compared with the optimal state $\ket{\phi}$ is defined as
\begin{equation}
    F = |\braket{\phi}{\psi}|^2.
\end{equation}

Finally, the \ac{nfev} are also tracked, which counts the number of times the objective function is evaluated on the quantum computer. It directly corresponds to the amount of resource usage of the quantum computer, and a lower value equates to a more resource-efficient algorithm.

\subsection{3-Uniform Cyclic Hypergraphs}

\begin{figure}[t]
  \centering
  \includegraphics[width=0.5\textwidth]{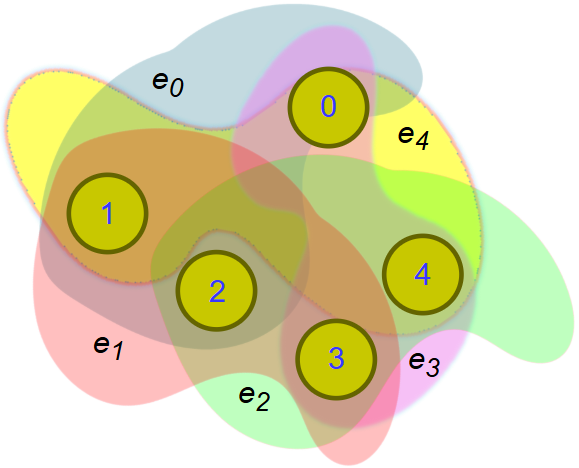}
  \caption{5-vertex 3-uniform cyclic hypergraph representation. Each node is represented by a yellow circle with a number $i$ inside it. The hyperedges are represented by areas, denoted by $e_i$, which encapsulate the nodes they link. For example, hyperedge $e_1$ links nodes $1$, $2$, and $3$.}
  \label{fig:5_cyclic_hypergraph}
\end{figure}

The $k$-uniform cyclic sign-alternating \ac{PUBO} Boolean problem can be expressed as follows
\begin{equation}
        C(\bm{x}) = \sum\limits_{i=0}^{n-k} (-1)^i\prod_{j=0}^{k-1} x_{i+j},
    \label{eq:XA_k_uniform_cyclic_hypergraph}
\end{equation}
where $x_{i + n} = x_i$. Specifically we investigate 3-uniform cyclic hypergraphs, which simplify to
\begin{equation}
    C(\bm{x}) = \sum\limits_{i=0}^{n-3} (-1)^i x_{i}x_{i+1}x_{i+2}.
\end{equation}
For example, for $n=5$, the associated hypergraph can be seen in Fig. \ref{fig:5_cyclic_hypergraph}.

Next, we transform the problem to a spin problem
\begin{align}
    C(\bm{z}) &= \frac{1}{8} \left[ \left( -\sum\limits_{i=0}^{n-k} (-1)^i \sigma^z_i \right) - 2(n \bmod 2) \sigma^z_1 \right. \nonumber \\
    &+ \left( \sum\limits_{i=0}^{n-k} (-1)^i \sigma^z_i \sigma^z_{i+2} \right) + 2(n \bmod 2) \sigma^z_0 \sigma^z_1 \nonumber \\
    &\left. - \left(\sum\limits_{i=0}^{n-k} (-1)^i \sigma^z_i \sigma^z_{i+1} \sigma^z_{i+2}\right) \right].
\end{align}
Therefore, the number of terms required for single-term interaction is $n$ (since the
$\sigma^z_1$ term is already present in the summation), $n + (n\bmod 2)$ terms are required for the two-term interaction, and $n$ terms are required for the three-term interaction, making the total of $3n + (n\bmod 2)$ terms required for the cost function. To this, we add $n$ for the number of terms in the mixer operator, for a total of $4n + n \bmod{2}$ parameters.

Figures \ref{fig:cyclic_AR}, \ref{fig:cyclic_AP}, and \ref{fig:cyclic_CC} in Appendix \ref{app:3-uniform_results} present the results for 3-uniform cyclic sign-alternating hypergraphs, which are pertinent instances to investigate given the computational hardness of related problems \cite{Guruswami2010}. Already at $p=1$ experiments, the \ac{AR} of the \ac{kA-QAOA} variant shows that it is an ideal candidate for 3-uniform cyclic sign-alternating hypergraphs. It gives better results than those achieved by \ac{AA-QAOA} and \ac{MA-QAOA} while providing a much better solution to \ac{SA-QAOA}. At $p=2$, \ac{AA-QAOA} and \ac{MA-QAOA} reach the same level of solution quality, while \ac{SA-QAOA} does not even reach a similar \ac{AR} or fidelity at $p=3$. Moreover, the \ac{nfev} required to reach the optimum by \ac{kA-QAOA} is considerably lower than that required by \ac{AA-QAOA} and \ac{MA-QAOA}. The notable drop in the \ac{nfev} highlights \ac{kA-QAOA}’s ability to deliver strong optimization performance with limited resource usage, positioning it as an effective approach for the constraints of the \ac{NISQ} era.

\subsection{Random Coefficient Hypergraphs}

To further evaluate the performance of \ac{kA-QAOA} against \ac{SA-QAOA} and \ac{MA-QAOA}, tests were performed on forty-four different 12-vertex hypergraphs, with randomly assigned coefficient values $J_i \in \{-1, 0, +1\}$, and $p \in [1, 5]$ with $k = 3$. Since only a few graphs exhibited automorphisms, \ac{AA-QAOA} was not tested. The cost function can be written as
\begin{equation}
    C(\bm{x}) = \sum_{i=0}^{n-1} J_i\prod_{j=0}^{2} x_{i + j},
    \label{eq:C3:3_uniform_randomC_hypergraph}
\end{equation}
where $x_{i + n} = x_i$.

Figure \ref{fig:random_coefficient} in Appendix \ref{app:random_coefficient_results} presents the results for the random coefficient hypergraphs, $J_i \in \{-1, +1, 0\}$ with $n = 12$. The \ac{kA-QAOA} variant starts to show a significant \ac{AR} at $p=2$, reaching the same levels of \ac{MA-QAOA} at $p=3$, while always providing a much better solution compared to \ac{SA-QAOA}. Moreover, as in the case of the previous experiments, the \ac{nfev} required to reach the optimum required is similarly lower than that needed by \ac{MA-QAOA}.

\subsection{Magic}

Following the work carried out by \citet{Capecci2025}, we wanted to investigate whether a magic barrier was witnessed when running \ac{MA-QAOA} and \ac{kA-QAOA} on hypergraphs, and compared the behavior with \ac{SA-QAOA}. We took 70 random 12-vertex hypergraphs consisting of randomly assigned integer coefficient values $J_i$, this time ranging between $-10$ and $+10$, up to five layers, where we evaluate the magic buildup throughout the circuit, specifically $M_2$. We also ensured that there were no ground state degeneracies in each instance to associate high-fidelity solutions with low magic. Here we only utilize the POWELL optimizer with \ac{TQA} parameter initialization.

\begin{figure}[t]
    \centering
    \subfloat[\label{fig:magic}]{%
        \includegraphics[width=\linewidth]{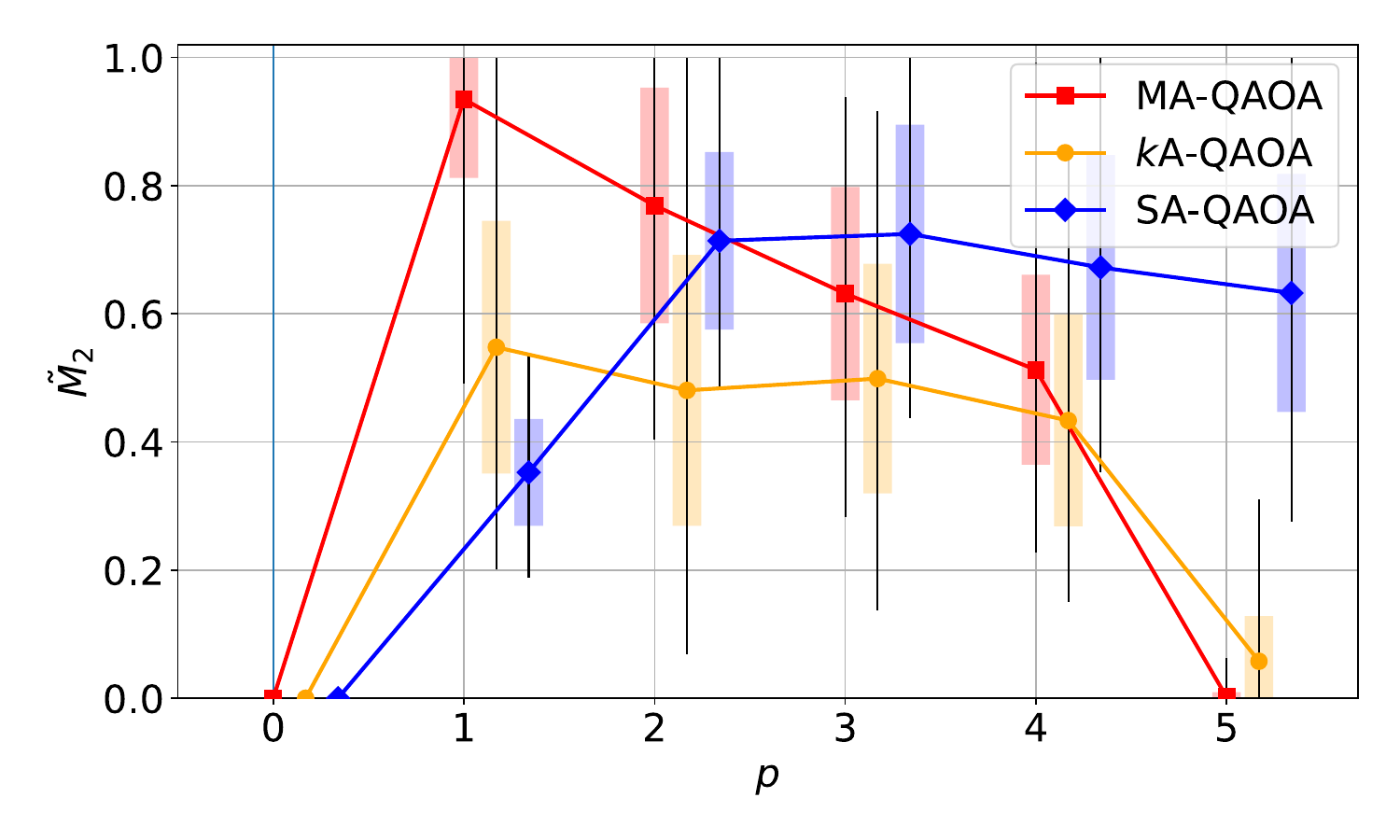}
    }\\
    \subfloat[\label{fig:AR}]{%
        \includegraphics[width=\linewidth]{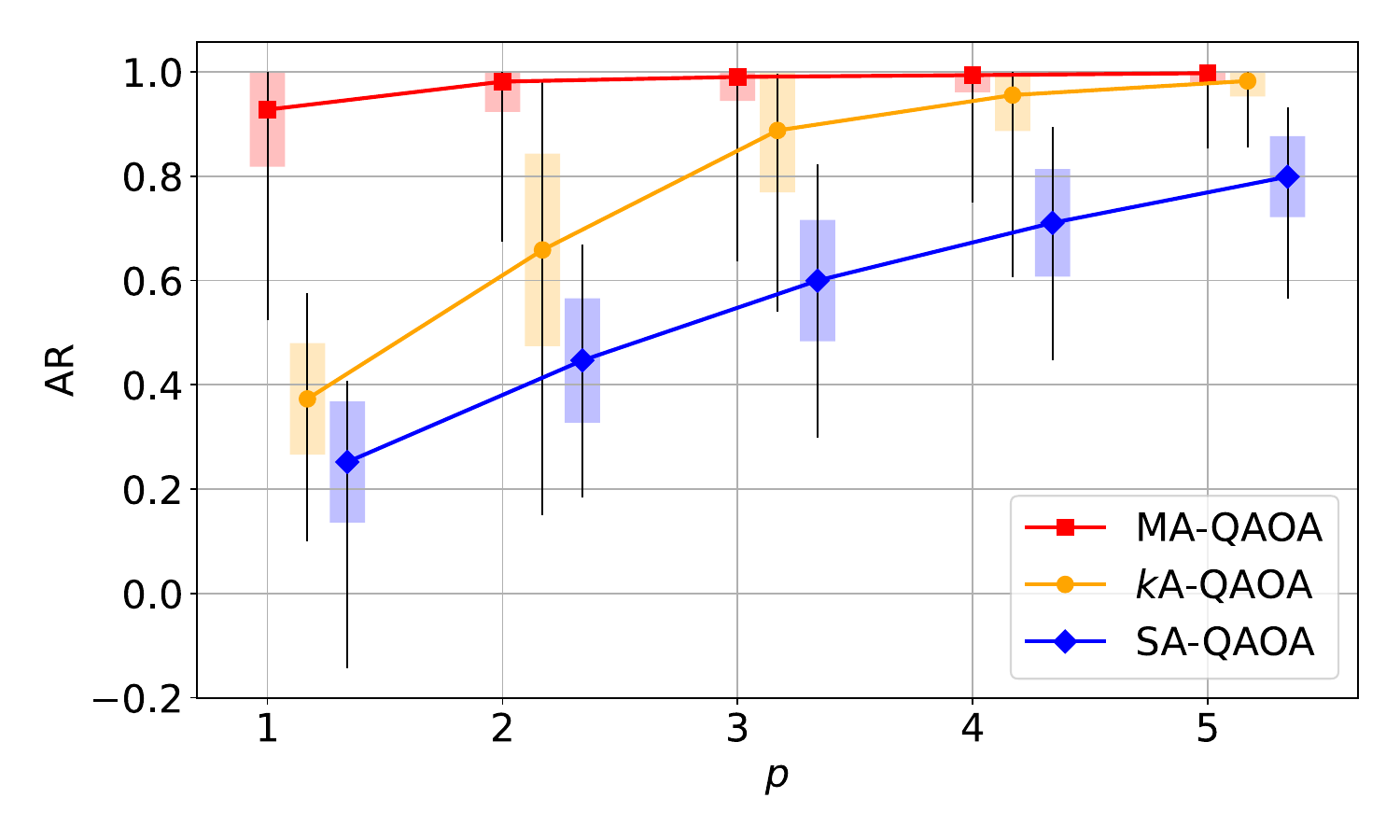}
    }\\
    \subfloat[\label{fig:nfev}]{%
        \includegraphics[width=\linewidth]{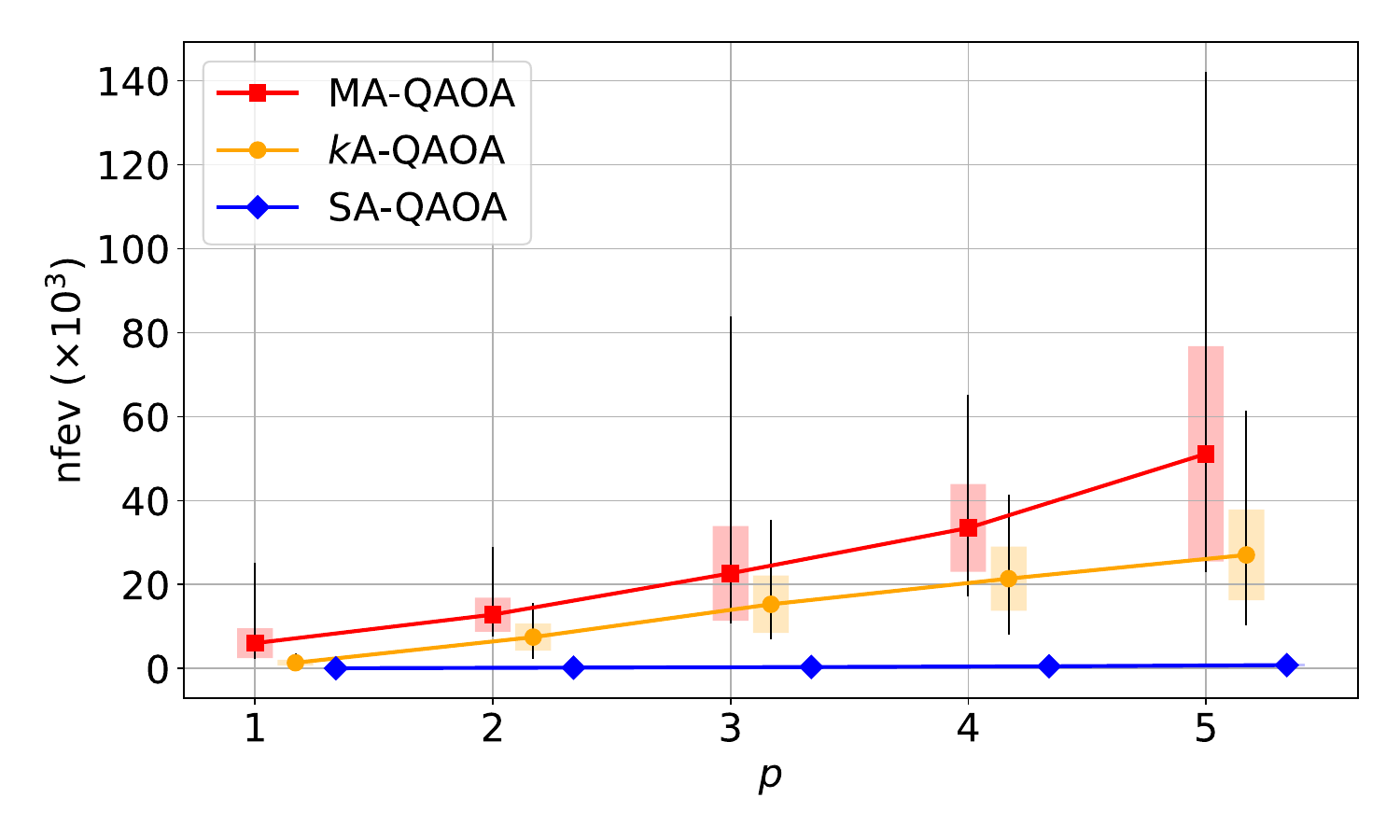}
    }
    \caption{Comparison of QAOA performance metrics: (a) average normalized magic, (b) average \ac{AR}, and (c) average \ac{nfev} for the POWELL optimizer with \ac{TQA} parameter initialization, all shown with standard deviation.}
    \label{fig:combined_vertical}
\end{figure}

Figure \ref{fig:magic} showcases the average and standard deviation of normalized magic $\tilde{M}_2$ as a function of the number of layers whilst Fig. \ref{fig:AR} shows the average and standard deviation of the \ac{AR} and Fig. \ref{fig:nfev} the \ac{nfev} for \ac{SA-QAOA}, \ac{kA-QAOA} and \ac{MA-QAOA} from one to five layers. Magic is normalized on a per-hypergraph basis by dividing each Rényi entropy value $M_2$ by the maximum entropy observed within that specific hypergraph. This maximum is computed across all layers and all parameterization strategies (MA, $k$A, and SA), keeping the highest value, which is scaled to 1, while all other values are expressed as relative proportions within the interval $[0, 1]$. This equates to rescaling each hypergraph instance by $1/\max_{p, \text{SA}, k\text{A}, \text{MA}} M_2$. Normalization is carried out to be able to compare relative magic between different hypergraphs and their potential solutions.

As expected, a magic barrier is present for all variants of \ac{QAOA}, and from Fig. \ref{fig:AR}, we find that the \ac{kA-QAOA} and \ac{MA-QAOA} converge to a solution after 5 layers, whereas the generic \ac{SA-QAOA} does not. We find that, on average, the \ac{kA-QAOA} magic barrier is lower than that \ac{MA-QAOA}, which in combination with a lower number of classical resources (reduced number of parameters and \ac{nfev}), also exhibits a reduction in quantum resources (magic). The caveat is that the average \ac{AR} for \ac{kA-QAOA} is slightly lower when compared with \ac{MA-QAOA}, although we contend that the resource-efficiency of \ac{kA-QAOA} compared with \ac{MA-QAOA} outweighs the negative. 

\section{Conclusion} \label{sec:Conclusion}

In summary, this work introduced the \ac{kA-QAOA}, a parameterization scheme that groups cost function terms according to their $k$-body interaction order to provide a natural balance between parameter efficiency and solution quality. We comprehensively benchmarked \ac{kA-QAOA} against standard \ac{SA-QAOA}, \ac{MA-QAOA}, and \ac{AA-QAOA} variants across two distinct problem classes: 3-uniform cyclic sign-alternating hypergraphs and random coefficient hypergraphs. Our results demonstrate that \ac{kA-QAOA} achieves approximation ratios comparable to the more expressive \ac{MA-QAOA} while requiring significantly fewer function evaluations. Furthermore, our investigation into nonstabilizerness revealed that \ac{kA-QAOA} traverses a lower average magic barrier during optimization compared to the highly expressive \ac{MA-QAOA}. We argue that this lower magic requirement is not merely an incidental observation but a significant advantage for \ac{NISQ}-era algorithms.

In many variational circuits, an excess of magic can lead to an over-parameterization of the quantum state without a corresponding increase in the \ac{AR} or fidelity, effectively wasting quantum resources. By grouping interactions by their $k$-body order, \ac{kA-QAOA} appears to selectively utilize non-stabilizer resources only where they are functionally relevant to the problem's topology. Consequently, \ac{kA-QAOA} provides a more resource-efficient path toward the global optimum, suggesting that it might be less susceptible to the noise accumulation that typically accompanies high-magic states in near-term hardware.

This resource efficiency, combined with its natural affinity for hypergraph structures, makes \ac{kA-QAOA} a promising candidate for solving real-world scheduling problems, such as the \ac{JSSP}. By reducing the number of variational parameters while maintaining high approximation performance, \ac{kA-QAOA} addresses one of the primary bottlenecks in scaling \ac{QAOA} to larger problem sizes on \ac{NISQ} devices. Future research could explore the performance of \ac{kA-QAOA} on other combinatorial optimization problems and investigate its synergy with hardware-native multi-qubit gates. Ultimately, the relationship between nonstabilizerness and optimization success warrants deeper investigation to guide the development of more efficient quantum-classical optimization loops.

\begin{acknowledgments}

M.C., T. J. G. A., and A.X. acknowledge funding from Project QAMALA (Quantum Algorithms and MAchine LeArning) financed by the Maltese Ministry for Education, Sport, Youth, Research, and Innovation (MEYR) Grant ``2025301 UM MinED''.

\end{acknowledgments}



\appendix

\onecolumngrid

\clearpage

\section{Graphs for 3-Uniform Cyclic Hypergraphs} \label{app:3-uniform_results}

\def\pvalues{1, 2, 3}

\foreach \gcode/\gtext in {AR/Approximation Ratio, AP/Fidelity, CC/\Acl{nfev}} {
  \begin{figure}[ht]
    \centering
    \foreach \p in \pvalues {
      \begin{subfigure}{\textwidth}
        \centering
        \includegraphics[width=\textwidth]{figures/plot_\gcode_gCyclic_p\p_xAB.png}
        \caption{\gtext\ for 3-Uniform Cyclic Sign-Alternating Hypergraphs with $p=\p$.}
        \label{fig:c4:sub:\p_\gcode_cyclic}
      \end{subfigure}
    }
    \caption{\gtext\ for 3-uniform cyclic sign-alternating hypergraphs with $p \in \{\pvalues\}$ for the cost function $C(\bm{x}) = \sum_{i=0}^{n-k} (-1)^i \prod_{j=0}^{k-1} x_{i+j}$ where $n \in [4, 15]$, $k=3$.}
    \medskip
    {\textit{\small Legend:} \usebox{\boxred} Multi-angle, \usebox{\orangecircle} $k$-interacting angle, \usebox{\bluediamond} Single-angle, \usebox{\greentriangle} Automorphic-angle.}
    \label{fig:cyclic_\gcode}
  \end{figure}

  \newpage
}

\clearpage

\section{Graphs for Random Coefficient Hypergraphs} \label{app:random_coefficient_results}

\begin{figure}[ht]
  \centering
  \foreach \gcode/\gtext in {AR/Approximation ratio, AP/Fidelity, CC/\Acl{nfev}} {
    \begin{subfigure}{\textwidth}
      \centering
      \includegraphics[width=0.5\textwidth]{figures/plot_\gcode_xC.png}
      \caption{\gtext}
      \label{fig:c4:sub:C12_\gcode_random}
    \end{subfigure}
    \\
  }
  
  \caption{Comparison of approximation ratio, fidelity, and \acl{nfev} for random coefficient $J_i \in \{-1, +1, 0\}$ hypergraphs with $p \in [1, 5]$ for the cost function $C(\bm{x}) = \sum_{i=0}^{n-1} J_i \prod_{j=0}^{k-1} x_{i+j}$ where $n=12$ and $k=3$.}
  
  \medskip
  {\textit{\small Legend:} \usebox{\boxred} Multi-angle, \usebox{\orangecircle} $k$-interacting angle, \usebox{\bluediamond} Single-angle.}
  \label{fig:random_coefficient}
\end{figure}

\clearpage

\twocolumngrid

\bibliography{ref}

\end{document}